\documentstyle[prl,twocolumn,aps]{revtex}
\begin{document}
\draft

\title{ Hidden Breit-Wigner distribution and other properties\\
of random matrices with preferential basis}

\author {Ph. Jacquod$^{(a)}$, D.L.Shepelyansky$^{(b,c)}$}
\address {$^{(a)}$ Institut de Physique, Universit\'e de Neuch\^atel,\\
1, Rue A.L. Breguet, 2000 Neuch\^atel, Suisse \\
$^{(b)}$ Laboratoire de Physique Quantique, Universit\'{e} Paul Sabatier,\\
118, route de Narbonne, 31062 Toulouse, France
}
%
%

\twocolumn[
\date{\today}
\maketitle
\widetext

\vspace*{-1.0truecm}

\begin{abstract}
\begin{center}
\parbox{14cm}{We study statistical properties of a class of
band random matrices which
naturally appears in systems of interacting particles.
The local spectral density is shown to follow the Breit-Wigner distribution
in both localized and delocalized regimes with width independent on the
band/system size. We analyse the implications of this distribution
to the inverse participation ratio, level spacing statistics and the problem of
two
interacting particles in a random potential.
}
\end{center}
\end{abstract}
\pacs{
\hspace{1.9cm}
}
]
\narrowtext
Intensive investigations of band random matrices (BRM) have been done
during last years \cite{casati,fyod}. Different regimes corresponding to
localized and delocalized wave-functions have been studied numerically and
analytically and it has been shown that the transition from one regime to
another can be described by one
scaling function depending on the ratio of the localization length in the
infinite system $l \sim b^{2}$
to the size of the matrix $N$, where the parameter $b$ determines
the size of the band $2 b+1$. Similar types of matrices appear in such
physical systems as quasi one-dimensional disordered wires and such models of
quantum chaos like the kicked rotator that gives additional grounds for
investigation of BRM.

The above BRM can be also considered as a reasonable model of one-particle
localization in a disordered wire of finite size \cite{casati,fyod}.
However recent investigations
of two interacting particles (TIP) in a random potential \cite{TIP} showed that
another type of BRM naturally appears in interacting systems. Indeed for
interacting particles there is one preferential basis which corresponds to
eigenstates without interaction. In this basis, the total
Hamiltonian is the sum of a diagonal matrix, with elements given by the sum of
one-particle energies, and a BRM, which describes interaction induced
transitions between eigenstates of the non-interacting problem. The first
investigations of such
superimposed BRM (SBRM) allowed to find the dependence of the localization
length $l_{sb}$ on the amplitude $W_{b}$ of large fluctuations on the diagonal
and to obtain the
localization length $l_{c}$ for two-particles coherent propagation in a
random potential on a distance much larger than one-particle localization
length $l_{1}$ \cite{TIP}.

While from the TIP model it is clear that matrices with preferential
basis should describe interesting physical effects in interacting systems, only
few investigations in this direction have been done up to now
\cite{haake,pich}.
In this paper we investigate the properties of such matrices in particular the
local spectral density and the inverse participation ratio (IPR).
Due to the close
connection between the SBRM and the TIP problem, the obtained results can
also be used for the latter case.

The matrix we study is the sum of a random diagonal matrix
and a conventional BRM :
\begin{equation}
 H_{n,n'} = \eta_{n} \delta_{n,n'} + \zeta_{n,n'}/\sqrt{2b+1}
\end{equation}
with $-W_{b} \leq \eta_{n}
\leq W_{b}$, $-1 \leq \zeta_{n,n'} \leq 1$ for $|n-n'| \leq b$ and
$\zeta_{n,n'} = 0$ elsewhere. The connection with the TIP
is given by $b \sim l_{1}^{2}$ and $W_{b} \sim 4 \sqrt{l_{1}}
V/U$ in terms of the
interaction strength $U$ and the one-particle energy bandwidth $4V$,
$l_1 \gg 1$.
This matrix describes a one-dimensional two-particle
Anderson model, with on-site interaction $U$, in the basis of non-interacting
eigenstates.
In \cite{TIP} it was shown that the eigenstates of (1) are localized
with localization length $l_{sb} \approx b^{2}/2 W_{b}^{2}$ for
$1 < W_{b} \ll \sqrt{b}$.
This leads to an enhancement of the length
of coherent TIP propagation $l_{c}=l_{sb}/l_{1} \sim l_{1}^{2}
\left(U/V\right)^2/32$ independent on the sign of interaction.

Our numerical investigations of SBRM (1) show that, in addition
to the standard exponentially localized form, the eigenstates are also
characterized by large amplitude fluctuations of probability on nearby sites.
A typical example of such an eigenstate is presented in Fig.1. The spike
eigenstate structure is clearly noticeable.

This implies that only certain unperturbed states have strong admixtures into
the given eigenstate.

Such eigenstate structure is quite different
from the case of conventional BRM. For a better understanding
of these spiked fluctuations
we study the local spectral density $\rho_{W}$
introduced by Wigner \cite{wign} and
analyzed in BRM with linearly growing diagonal corresponding to conservative
systems \cite{boris,sydney} :
\begin{equation}
\rho_{W}(E-E_{n}) = \sum_{\lambda} \vert \psi_{\lambda}(n) \vert^{2}
\delta(E-E_{\lambda})
\end{equation}
The function $\rho_{W}$ characterizes the average probability
$P(\vert \psi_{\lambda}(n)
\vert^{2}) = \rho_{W}(E-E_{n})$ of eigenfunction
$\psi_{\lambda}(n)$ on site $n$ with energy $E_{n}=H_{n,n}$, where $\lambda$ is
the eigenvalue index and $n$ marks the original basis.
Our numerical investigations
in a wide range of parameters ($20 \leq b \leq 2000$, $201 \leq N \leq 4001$
and $ 1.5 \leq W_{b} \leq 40 $) both in localized ($l_{sb} \ll N$)
and delocalized
($l_{sb} \gg N$) regimes show (see Fig.2)
that $\rho_{W}$ is well described by the well-known
Breit-Wigner distribution $\rho_W = \rho_{BW}$:
\begin{equation}
\rho_{BW}(E-E_{n})=
\frac{\Gamma}{2 \pi((E-E_{n})^{2}+\Gamma^{2}/4)}; \\
\Gamma = \frac{\pi}{3 W_{b}}
\end{equation}
where $\Gamma$ is the distribution width.
This distribution remains valid
in localized and delocalized regimes under the condition that
$\Gamma$ is much less than
the energy width $\delta E \approx 1$ of $H_{n,n'}$ at $W_{b}=0$.
Usually the distribution
$\rho_{BW}$ appears in such physical systems as nuclei
and complex atoms \cite{sydney}
where due to energy conservation the diagonal term $\eta_{n}$ grows linearly
with $n$ that corresponds to a finite level density $\rho_{E}$.
In this case the width is
$\Gamma = 2 \pi \rho_{E} <H_{n \neq n'}^{2}>$
\cite{wign,sydney}.
In our case for $W_{b} \gg 1$ all eigenenergies are homogeneously distributed
in the finite interval $[-W_{b},W_{b}]$ and for full matrices with $b=N/2$
we can use the above expression with $\rho_{E}=N/2W_{b}$
which gives $\Gamma$ in (3). For $b \ll N$
according to \cite{TIP} one should replace $\rho_{E}$
by the density of directly
coupled states $\rho_c=b/W_{b}$ that leads to the same expression for $\Gamma$.
The
theoretical formula for $\Gamma$, independent on $b$ and $N$,
is in a good agreement with our numerical data (Fig.2). The independence of
$\Gamma$ on $b$ and $N$ makes our case quite different from the case
of full matrix (1) studied before in \cite{tat}.

For $W_{b} \gg 1$ the width of the Breit-Wigner
peak is small and therefore according to (3)
the probability on nearby levels is a strongly fluctuating spiked function.
This spike structure of eigenfunctions can be characterized
by the IPR $\xi_{\lambda} =(\sum_{n} |\psi_{\lambda}(n)|^{4})^{-1}$
which counts
the number of spikes independently on the distance between them. In the case of
full
matrices with $b=N/2$ the number of spikes
can be estimated as the number of states
in the interval $\Gamma$ that gives the average value of IPR
$\xi = <1/\xi_{\lambda}>^{-1} \sim \rho_{E} \Gamma \approx N/(2 W_{b}^{2})$.
The same estimate can be also used in the delocalized regime $l_{sb} \gg N$
with $b \ll N$. Of course
this estimate is valid only when the number of states in the width $\Gamma$
is much larger than one, that implies $\xi \gg 1$ or $W_{b} \ll \sqrt{N}$.

The numerical results for the dependence of IPR on $W_{b}$ in the delocalized
regime are presented in Fig.3. They demonstrate that for sufficiently large
full matrices ($N=4001$) this dependence approaches to the above estimate.
However the convergence is rather slow so that for smaller $N$ values
one has approximately $\xi \sim N/W_{b}^{\alpha}$ where the exponent $\alpha$
slowly changes with $N$. For example $\alpha \approx 1.7$ for $N=2001$. We
attribute this very slow approach to the asymptotic value of $\alpha=2$ to
the quite restricted range of $W_{b}$ variation.
Indeed on one side
the width of the Breit-Wigner peak should not exceed the width of the
energy band for $W_{b}=0$ that gives $W_{b} \gg 1$. On the other side one
should have $W_{b} \ll \sqrt{N}$. Another restriction appears for band
matrices with $b < N/2$ namely $l_{sb} \gg N$. The data for this case
(Fig.3, full squares) show that for not very large $W_{b}$
the IPR is close to the regime of
full matrices while for large $W_{b}$ one enters the localized regime
$l_{sb} \ll N$ which should be studied separately.

It is interesting to note that in the delocalized regime even for
$W_{b} \gg 1$ many levels are coupled by interaction
if $\rho_{E} \Gamma \approx N/(2 {W_{b}}^2) \gg 1$.

Therefore, one would
expect that for $W_b < {W_b}^{cr} \approx (N/2)^{1/2}$
the level spacing statistics $P(s)$ will be the same as
in the case of Gaussian orthogonal ensemble (GOE) \cite{metha}.
These expectations are not so evident since the spiked
structure of eigenfunctions apparently should
lead to a decrease of overlapping matrix elements between
eigenfunctions. However, our numerical results for
matrices with $N \leq 8000$ show that $P(s)$ remains close to GOE
for $1 < W_b < {W_b}^{cr}$. They are also in agreement with
the numerical results \cite{lenz} for full matrices of smaller sizes
showing that the transition
border in $W_b$ between Poisson and GOE statistics scales
as $N^{1/2}$. The question about other statistical properties
of levels in the regime $1 < W_b < {W_b}^{cr}$ remains open.

For the localized regime in the above estimate of $\xi$ one should
replace $N$ by $l_{sb}$ since only levels in the interval of
one localization length can contribute to the IPR. This gives the expression
\begin{equation}
\xi \approx l_{sb}/2{W_b}^{\beta-2} \approx b^{2}/4 W_{b}^{\beta};
\; \; \; \beta=4
\end{equation}
which is valid for $\xi \gg 1$
($1 \ll W_{b} \ll \sqrt{b}$). The last condition together with
$l_{sb}\ll N$ gives strong restrictions for the numerical simulations
($1 \ll W_{b} \ll N^{1/4}$).

Our results for this localized case are presented in Fig.4.
The data can be empirically fitted by $\xi \sim b^{2}/W_{b}^{\beta}$
with $\beta \approx 3$ which differs from the theoretical value
$\beta = 4$. We attribute this difference to the fact that we are not
far enough in the asymptotic regime of large $b$ and $W_{b}$.
Indeed, for $W_{b} > b^{1/2}$ one enters in the perturbative
regime and the deviations from a power law becomes evident.
We also checked that the probability distribution
$P(\vert \psi_{\lambda}(n) \vert^4)$ is proportional to
$\rho_{BW}^2$ that gives additional grounds for the theoretical
power $\beta=4$. However, the simulations
with large enough values of parameters $b, W_{b}$ requires
too large matrix sizes
being beyond our numerical abilities. The numerically found value $\beta > 2$
implies that the number of peaks is smaller than the localization
length $l_{sb} \approx b^{2}/(2 W_{b}^{2})$ which determines the
asymptotic exponential decay of the eigenstates. It would be desirable
to have a more rigorous theoretical derivation of the IPR dependence on
parameters in the localized regime.

The above results show that the SBRM (1) has many features similar
with the photonic localization in a molecular quasicontinuum \cite{PHD}
as it was remarked in \cite{TIP}. According to this analogy,
the number of levels in one-photon transition
(size) is of the order $b$ and the density of coupled states
is $b/2W_b$. However, in the photonic model
the levels are ordered in energy in a growing way that leads to a chain of
equidistant Breit-Wigner peaks in an eigenstate \cite{PHD}. For the SBRM (1)
all levels are mixed in the energetic interval and the Breit-Wigner
peak is hidden.

Let us now discuss the consequences of the result (4) for the TIP
model. According to the relation between the parameters of SBRM and TIP
given above we obtain from (4) the expression for the IPR $\xi_c$
in the TIP model:
\begin{equation}
\xi_{c} \sim (U/V)^4 {l_1}^2 > 1
\end{equation}
This result can be also derived directly from the
density of states inside the localization
length interval $l_c$ ($\rho_E \sim l_1 l_c/V$) and the
transition rate $\Gamma_c \sim U^2/(V l_1)$ obtained in \cite{TIP}.
Indeed, the number of levels in the Breit-Wigner peak is
$\Gamma_c \rho_E \approx \xi_c$ that gives (5). This result shows that the
number of noninteracting eigenstates $\xi_c$ contributing in the eigenfuction
is quite large for $U \sim V$ while for $(U/V)^2 l_1 \ll 1$
this number is order of 1. However, the value of $\xi_c$
at $U \sim V$ is much less than
the number of unperturbed states $\Delta N$ contributing
to the TIP eigenfunction in the unperturbed
lattice basis. This number
determines the IPR $\xi_{max} \approx \Delta N \approx l_c l_1 \sim l_1^3$.
The difference between $\xi_c$ and $\xi_{max}$ shows that
the noninteracting eigenbasis represents the real eigenfunctions
in a much better way. It also stresses the fact that
the IPR value is not basis invariant.

{}From the difference between $\xi_c$
and $\xi_{max}$ it is possible to conclude that the coherent
propagation of TIP goes by rare jumps of size $l_1$
between the states with
approximately constant sum of noninteracting energies
$E_s=\epsilon_n+\epsilon_{n'}$. The distribution over $E_s$
should have the Breit-Wigner form with the width $\Gamma_c$.
The length of propagation by such jumps is $l_c \sim {l_1}^2 \gg l_1$.
Due to this hidden Breit-Wigner distribution the IPR $\xi_c$
in the basis of noninteracting eigenstates is proportional
to ${l_1}^2$ instead of "naive" ${l_1}^3$. For the case
of TIP with $M$ transverse channels one should replace $l_1$
in (5) by $M l_1$ with $l_1 \propto M$ being one-particle localization length.

If one-particle motion is ergodic in a $d$-dimensional
system of size $L < l_1$ then its
eigenfunction contains about $N_1 \approx L^d$
components. The matrix element of interaction is then $U_s \sim U/{N_1}^{3/2}$
\cite{TIP}, the density of coupled states $\rho_c \sim {N_1}^2/V$
and the Breit-Wigner width $\Gamma_c \sim {U_s}^2 \rho_c \sim U^2/N_1 V$
for $U < V$ is less than one-particle level spacing $\Delta_1 \approx V/N_1$.
Thefore, it is possible that a concept of pairs formed by TIP can
be useful even in the ergodic samples with $L<l_1$.
In some sense, for $\Delta_2 \ll \Gamma_c \ll \Delta_1$,
where $\Delta_2 \approx V/{N_1}^2$ is two-particle level spacing,
one can at first average over fast one-particle motion
and after that analyse the slow pair dynamics with
typical time scale $1/\Gamma_c$.
In the ergodic regime $L < l_1$ the IPR is $\xi_c \sim \Gamma_c \rho_c
\sim N_1 (U/V)^2 \ll {N_1}^2$ and
according to the discussed above properties of $P(s)$ in
SBRM and the result \cite{lenz} the GOE statistics for TIP
should be observed for $\xi_c > 1$.
For $L \gg l_1$ the strong enhancement of interaction
($\rho_c \Gamma_c \sim {l_1}^d (U/V)^2 \gg 1$) leads to delocalization of
the TIP pairs in $d \geq 3$ below one-particle
Anderson transition when noninteracting particles are well localized
\cite{3d}.

One of us (P.J.) gratefully acknowledges the hospitality of
the Laboratoire de Physique
Quantique, Universit\'e Paul Sabatier during his stay there and another
(D.L.S.) is grateful to the Yale University and
the Godfrey Fund at University of New South Wales for hospitality
during the process of work on the above problem.
We thanks Y.Alhassid, V.Flambaum and O.Sushkov for
useful discussions and remarks. Work is supported in part by the
Fonds National Suisse de la Recherche.


\begin{figure}
\caption{Localized eigenfunction of a SBRM with $W_{b}=7$, $b=100$ and
$N=2001$.
The solid line indicates the exponential localization with $l_{sb} \approx 171$
in agreement with results obtained in [3], eq.(3).}
\label{fig1}
\end{figure}


\begin{figure}
\caption{Local spectral density which determines the average probability
on a given site $P(\vert \psi_{\lambda}(n) \vert^{2}) = \rho_{W}(E-E_{n})$
for $b=100$, $W_{b}=5$,
$N=201$ (triangles, 20 realisations of disorder)
and $N=2001$ (squares, 2 realisations of disorder).
The solid line gives Breit-Wigner distribution (3) with $\Gamma=0.21$.
The inset shows the dependence of $\Gamma$ on $W_{b}^{-1}$ : points are
numerical
data ($N=1001, b=100$), straight line is theory (3).}
\label{fig2}
\end{figure}


\begin{figure}
\caption{IPR $\xi$ normalized with its limit value
for the GOE case $N/3$ vs. $W_{b}$ in the delocalized regime for $N=2001$,
$b=300$ (full squares), and full matrices
with $N=251$ (open squares), $N=501$ (open triangles), $N=1001$ ($\times$),
$N=2001$ (full triangles) and $N=4001$ (full circles). Dashed line shows the
fit for full circles with $\alpha = 1.75 \pm 0.03$;
solid lines shows theoretical slope $\alpha=2$.}
\label{fig3}
 \end{figure}


\begin{figure}
\caption{Dependence of $\xi/b^{2}$ on $W_{b}$ in localized regime :
$N=2001$, $b=50$ (full squares) and $b=80$ (open squares); $N=4001$,
$b=50$ (full circles) and $b=100$ (open circles). Dashed line
shows the slope from fit for open circles ($\beta=3.0 \pm 0.1$)
and solid line indicates theoretical slope $\beta=4$.}
\label{fig4}
 \end{figure}
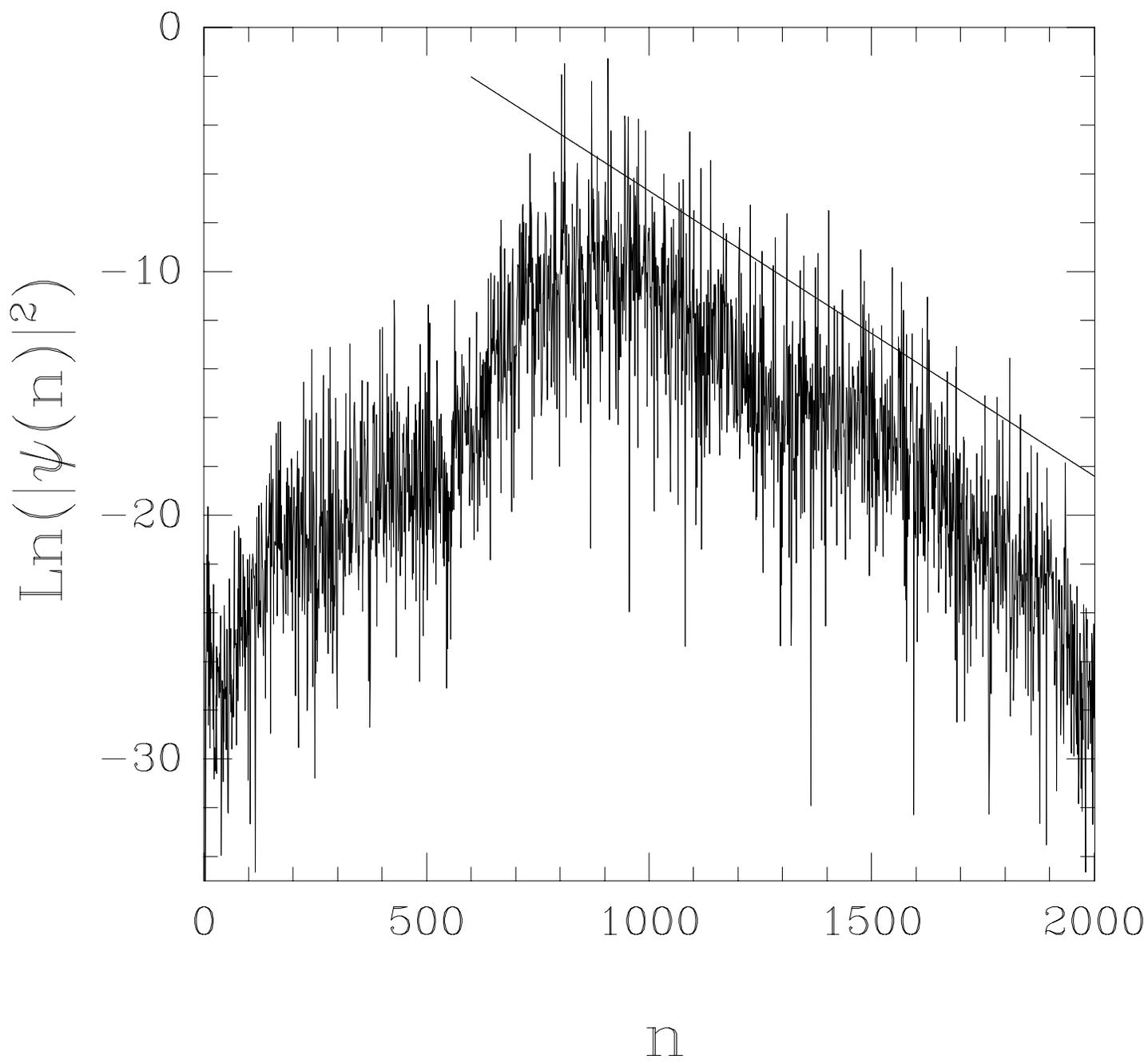
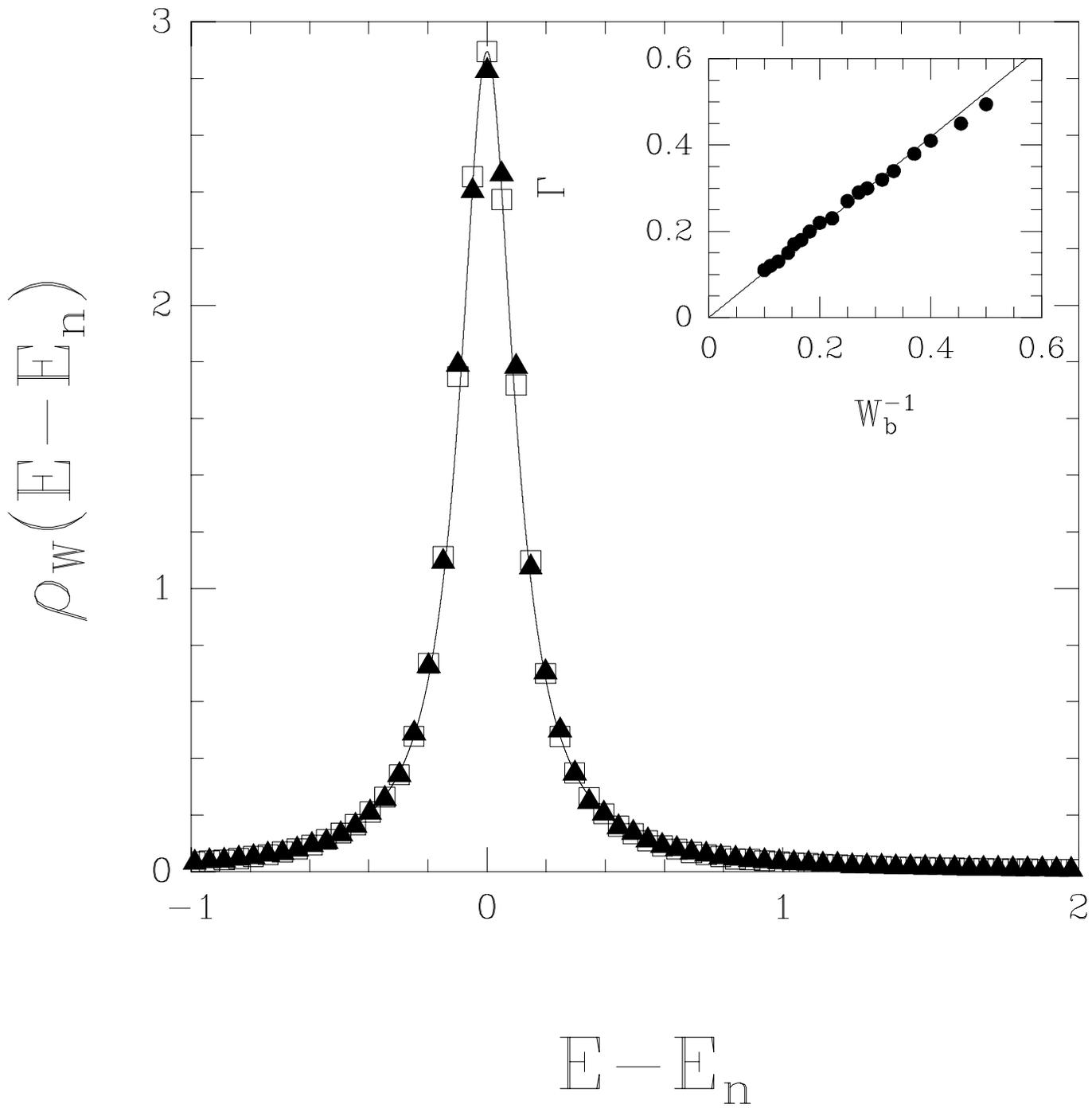
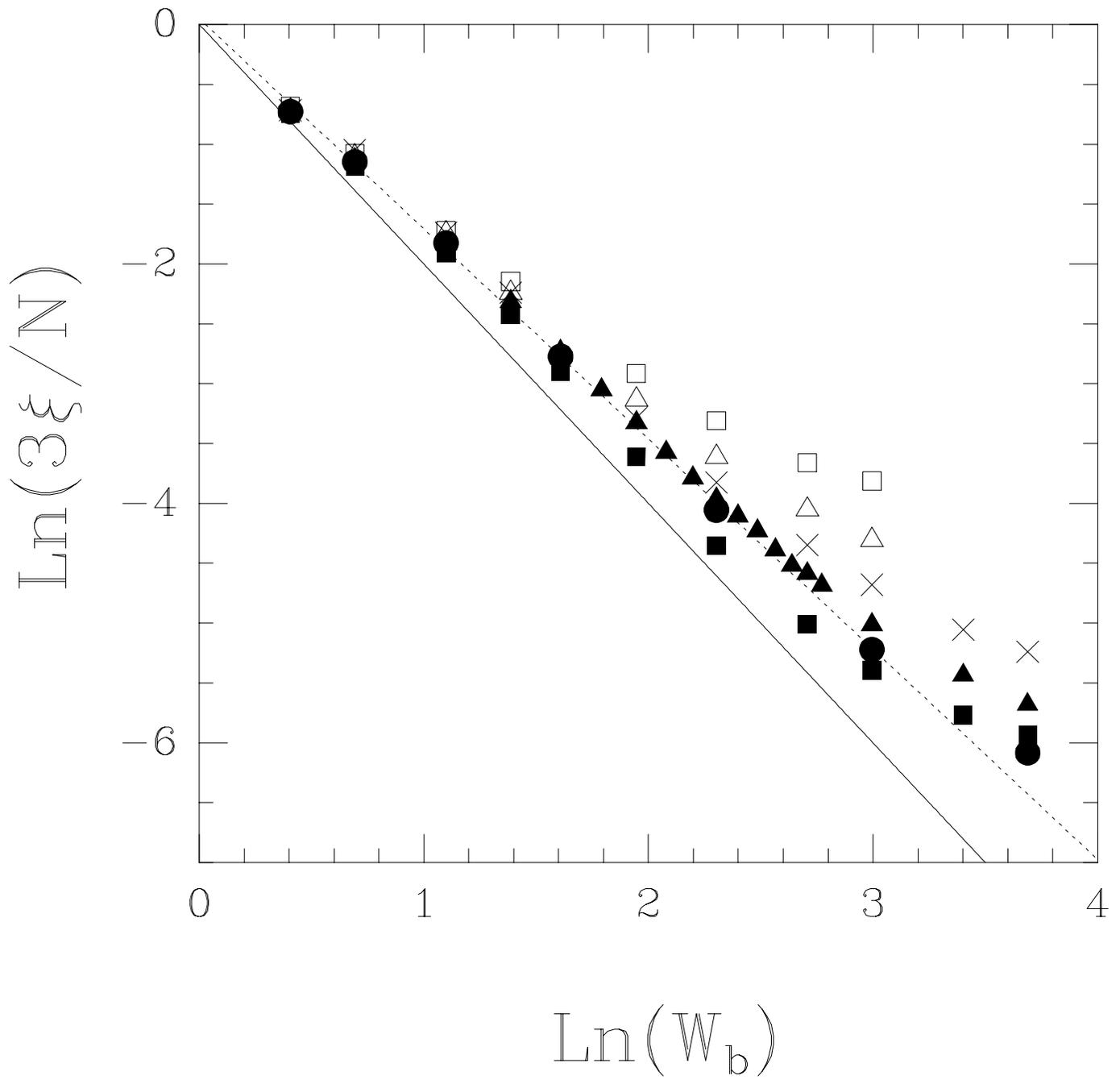
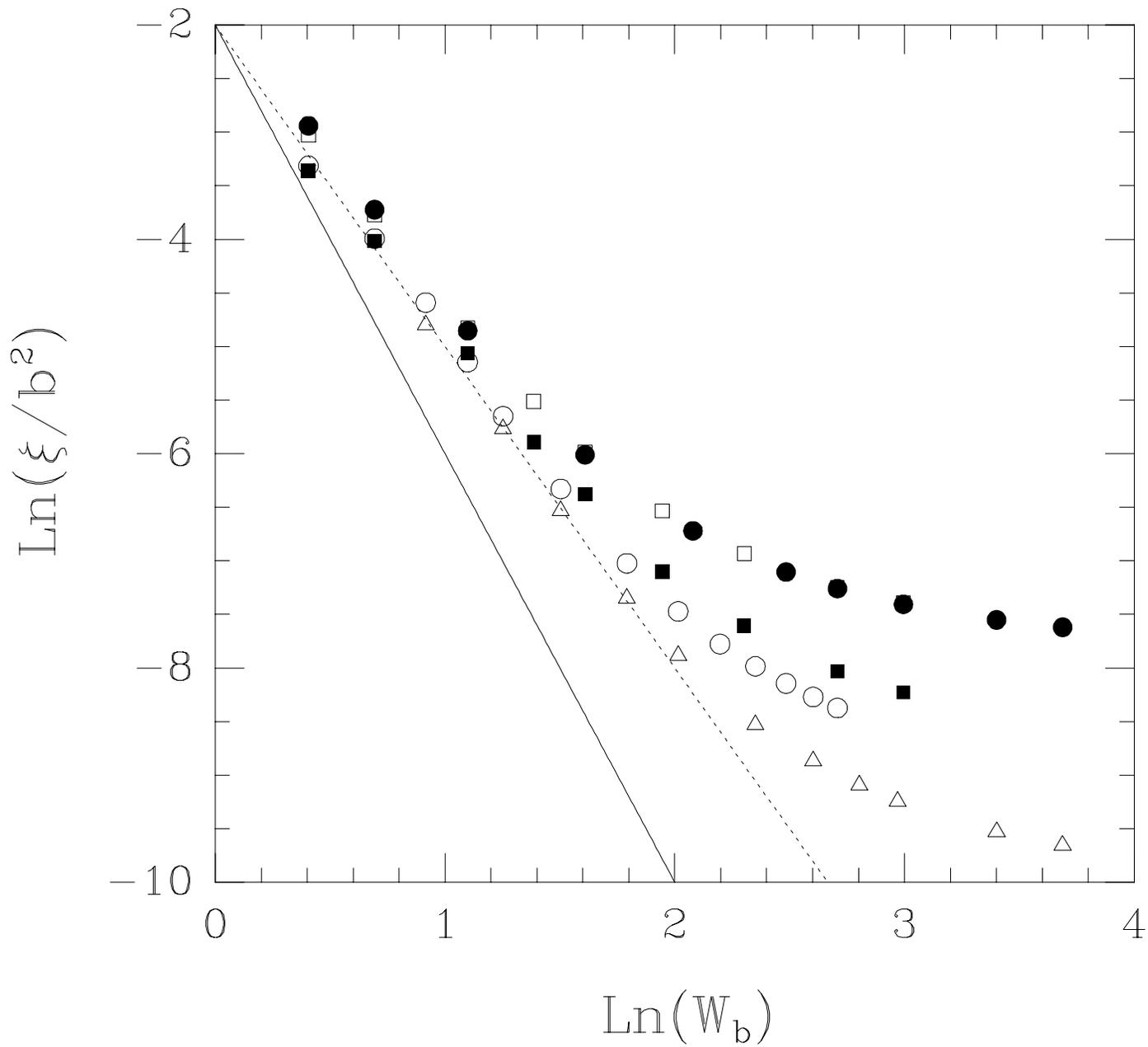
\end{document}